\documentclass[%
 aip,
 amsmath,amssymb,
reprint,
]{revtex4-1}

\usepackage{graphicx}
\usepackage{subfig}
\usepackage{dcolumn}
\usepackage{bm}

\usepackage[utf8]{inputenc}
\usepackage[T1]{fontenc}
\usepackage{mathptmx}
\usepackage{etoolbox}
\usepackage{xcolor}

\newcommand{\Ea}{\mathrm{E_a}}
\newcommand{\Echi}{\mathrm{E_\chi}}

\makeatletter
\def\@email#1#2{%
 \endgroup
 \patchcmd{\titleblock@produce}
  {\frontmatter@RRAPformat}
  {\frontmatter@RRAPformat{\produce@RRAP{*#1\href{mailto:#2}{#2}}}\frontmatter@RRAPformat}
  {}{}
}%
\makeatother
\begin{document}


\title[TCE model for surface processes in the DSMC method]{Total Collision Energy model for surface processes in the DSMC method}
\author{Yevgeniy A. Bondar}
 \affiliation{Khristianovich Institute of Theoretical and Applied Mechanics SB RAS, 4/1 Institutskaya str., Novosibirsk 630090, Russia}
\author{Alexandra N. Molchanova}
 \email{molchanova@itam.nsc.ru}
 \affiliation{Khristianovich Institute of Theoretical and Applied Mechanics SB RAS, 4/1 Institutskaya str., Novosibirsk 630090, Russia}

\begin{abstract}
The total collision energy (TCE) approach originally developed for taking into account gas-phase chemistry in the direct simulation Monte Carlo method is applied to surface chemistry modeling. In the proposed surface TCE (sTCE) model probability of the impact process as a function of the translational energy of the molecule is found analytically from the finite-rate Arrhenius data. This probability function lacks some mathematical drawbacks typical of previous models. Effects of surface temperature, incident angle and internal energy can be taken into account in the sTCE model. Additionally the model is generalized to the specified reaction efficiency formulation.  
\end{abstract}

\maketitle

The direct simulation Monte Carlo (DSMC) method~\cite{Bird-1994} is currently the principal numerical technique for modeling the gas flows with a high degree of thermal and chemical non-equilibrium. High-altitude aerothermodynamics of space vehicles is traditionally one of the main areas of DSMC applications~\cite{Bird-RGD-1979}. Modeling of various surface chemical and physical processes within the DSMC simulation is critical for accurate prediction of heat flux to the surface  of descent vehicles at high altitudes and even their aerodynamic characteristics~\cite{Litvintsev-FluidDyn-2024}.  

While modern methods of computer chemistry can be employed for obtaining accurate data for DSMC modeling such processes (see, e.g., works~\cite{Marschall-ChapterInBook-2015, Kroupnov-ActaAstronautica-2023, Xinbin-POF-2025}), these data normally cannot be directly used in the DSMC calculations and are not available for each elementary step of complex surface chemical mechanisms.
For this reason in the engineering applications simpler general models are required which can transform available computer simulation and experimental data into the form suitable for the modeling of surface chemistry in the DSMC.  

The SMILE++ software system~\cite{SMILE++RGD27} was the first major DSMC code to include a detailed framework for modeling surface chemistry~\cite{Molchanova-RGD29-2014, Molchanova-POF-2018}. The idea~\cite{Shumakova-AIAA-2014} was to obtain data for the DSMC surface chemistry modeling from finite-rate macroscopic mechanisms as it is traditionally done for the DSMC gas-phase chemistry~\cite{Bird-1994}. For a part of the elementary processes which involve a molecule which impinges on a surface  from the gas phase (so-called impact processes) the approach~\cite{Choquet-1999} was employed with some modifications which allowed obtaining the process probability in the analytical form as a function of the molecular velocity component normal to the surface. This model has been found to have substantial drawbacks, particularly, often the probability function is unlimited in the vicinity of the reaction threshold which leads to numerical errors~\cite{Molchanova-POF-2018, Molchanova-Lobachevskiy-2026}.   

The present work is devoted to overcoming this obstacle by applying the total collision energy (TCE) approach~\cite{Bird-RGD-1979, Bird-AIAA-1981} by G. Bird, which is probably the most popular in the DSMC community for the gas-phase chemistry~\cite{Gimelshein-BirdLegacy-2019}, to the impact surface processes. In the present paper, first, the assumptions of the original TCE model are given, then the surface TCE (or sTCE) model is presented for atomic species and compared with the approach~\cite{Choquet-1999, Molchanova-POF-2018}. Finally, several generalizations of the sTCE models are proposed in order to include the effects of surface temperature, angle of incidence and internal degrees of freedom of the impinging molecule. A variant of the sTCE model for the specified reaction efficiency approach is also presented.

\bigskip

\noindent{\bf   TCE model for gas phase reactions}

\smallskip

\noindent
The original TCE model will be briefly presented below. Let us consider a bimolecular gas phase 
reaction 
\begin{equation}
A+B \longrightarrow C + D
\label{eq:reaction}
\end{equation}
Number of reactions  per unit time and volume in thermal equilibrium conditions equals 
\begin{equation}
K \left( T \right)n_An_B,
\end{equation}
where   
 $n_A$ and $n_B$ are number densities of the reactants, $T$ is gas temperature, and $K(T)$ is the reaction rate constant given in the modified Arrhenius form
\begin{equation} 
 A T^b \exp \left( -\frac{E_a}{kT} \right), 
\label{eq:arrenius}
\end{equation}
where $k$ is the Boltzmann constant and $A$, $b$ and $E_a$ are known parameters. In the TCE model it is assumed that~\cite{Bird-AIAA-1981, Gimelshein-BirdLegacy-2019}:

{\bf 1.} the reaction occurs during the collision between molecules $A$ and $B$ with a probability $P$ (acceptance-rejection procedure is used); $P$ is a function of the total collision energy $E$ -- typically, a sum of rotational/vibrational energies of both molecules and their relative translational energy; 

{\bf  2.} the probability function $P(E)$ sought in the analytical form is such that it provides the reaction rate constant $K(T)$ in the form~\eqref{eq:arrenius} under thermal equilibrium conditions, so the constants $A$, $b$, and $E_a$ are considered as input parameters of the model; 

{\bf  3.} if the reaction occurs in the collision, then the reaction energy (which can be negative or positive) is added to the total collisional energy, which is then redistributed between the relative translational energy and rotational/vibrational modes of the reaction products; normally Larsen-Borgnakke procedure~\cite{LB-1975} is employed.

The function $P(E)$ is derived analytically in the conventional variant of the TCE model which assumes continuous rotational and vibrational energies and variable hard sphere~\cite{Bird-AIAA-1981} (or variable soft sphere~\cite{Koura-PhysFluids-1991}) model for elastic collisions.  The equilibrium Boltzmann distribution function of total collision energy $E$ of two molecules undergoing a binary collision  is given by 
\begin{equation} 
f(E,T)
= \frac{E^{\frac{\xi}{2}-1}}{\Gamma\!(\xi/{2})(kT)^{\xi/2}}
  \exp\!\left(-\frac{E}{kT}\right),
\label{eq:dfTCE}
\end{equation}
where $\xi$ is a total number of degrees of freedom of energy modes which contribute to the reaction and $\Gamma$ is the gamma function.  Then in thermal equilibrium the reaction  rate constant can be presented as follows 
\begin{equation} 
K(T)=g(T)\int\limits_{E_a}^{\infty}P(E)E^{\frac{\xi}{2}-1}\exp\!\left(-\frac{E}{kT}\right)dE,
\label{eq:kTCE}
\end{equation}
where $g(T)$ is a known power function of temperature which depends on parameters of the elastic collision model and masses of the reactant molecules. If $K(T)$ is known this expression can be considered an integral equation which can be solved to obtain the unknown function $P(E)$. By choosing the following form of the probability function
\begin{equation}
P \left( E \right) =C  \dfrac{{\left( E - E_a \right)}^{\alpha}}{E^{\frac{\xi}{2}-1}} \mathcal{H} \left( E - E_a \right),
\label{eq:pTCE}
\end{equation}
where $C$ and $\alpha$ are constants to be determined
and $\mathcal{H}$ is the Heaviside step function 
\begin{equation}  
\mathcal{H} \left( t \right) = \left\{
\begin{array}{l}
1, \,\,\,\,\,\,\,\,\,\,\,\, t \geq 0 \\ 
0 , \,\,\,\,\,\,\,\,\,\,\,\, t < 0
\end{array}
\right.
\label{eq:H}
\end{equation}
one gets rid of the $E^{\frac{\xi}{2}-1}$ factor in Eq.~\eqref{eq:kTCE} and the integral is reduced to the gamma function and can be analytically calculated. Eq.~\eqref{eq:kTCE} then takes the following form
\begin{equation} 
K \left( T \right) = C \beta_1 (kT)^{\alpha+\beta_2} \Gamma({\alpha} + 1) \exp \left( -\frac{E_a}{kT} \right), 
\label{eq:kTCE1}
\end{equation}
where $\beta_1$ and $\beta_2$ are known constants. The parameters $C$ and $\alpha$ are obtained analytically by equating the right part of Eq.~\eqref{eq:kTCE1} to the known $K \left( T \right)$ in the Arrhenius form~\eqref{eq:arrenius}. After that the probability function~\eqref{eq:pTCE} is fully determined and can be used in the DSMC calculations.

\bigskip

\noindent{\bf   General assumptions of the sTCE model}

\smallskip

\noindent
Let us now consider various physical and chemical processes occurring when a particle (usually neutral or charged atom or molecule)  impinges on a solid surface. 
These impact processes include but are not limited to adsorption 
\begin{equation}
\mathrm{A}_g + \, _S \longrightarrow \mathrm{A}_S \, ,
\label{eq:adsorption}
\end{equation}
and Eley-Rideal recombination
\begin{equation}
\mathrm{A}_g + \mathrm{B}_S \longrightarrow \mathrm{AB}_g + \, _S \, ,
\label{eq:ERrecombination}
\end{equation}
where $\mathrm{A}$ and $\mathrm{B}$ are particles, $\mathrm{AB}$ --- the product of their recombination, $_g$ stands for the gas phase, and  $_S$ denotes particles adsorbed by the surface on vacant adsorption sites. All these processes can be included as elementary steps in the detailed finite-rate surface chemistry mechanisms (models).
Note that the processes which include only particles adsorbed by the surface (e.g. Langmuir-Hinshelwood recombination or desorption) are not considered in the present work. 

At the macroscopic level number of reactions per unit time and surface area due any of these processes is given by  
\begin{equation}
K \left( T \right)n_{gas}n_{surf},
\label{eq:number}
\end{equation}
where $n_{gas}$ is number density of gas phase particles and $n_{surf}$ is a surface number density of adsorbed particles or vacant sites, and the reaction rate constant usually has the Arrhenius form~\eqref{eq:arrenius}.

The following assumptions are used in the surface TCE model for modeling such processes in the DSMC method:

{\bf 1.} the reaction occurs during the interaction ("collision") between a particle $A_g$ or $AB_g$ and a surface with a probability $P$ (acceptance-rejection procedure is used); $P$ is a function of the total collision energy $E$ (typically a sum of energies of all modes of the incoming particle) as well as various macroscopic parameters such as $n_{surf}$ and the surface temperature $T$;   

{\bf 2.} the probability function $P(E)$ sought in the analytical form is such that it provides the Arrhenius dependence of the reaction rate constant $K(T)$ under thermal equilibrium conditions with equal temperature $T$ of both gas and surface, so the constants $A$, $b$, and $E_a$ are considered as input parameters of the model; 

{\bf 3.} if the reaction occurs, then the reaction energy (which can be negative or positive) is added to the total collisional energy; the energy then is redistributed between the heat transferred to the surface and energy modes of the reaction product particles emitted to the gas phase according to a prescribed energy redistribution rule.

Note that the rules of energy redistribution is beyond the scope of the present work. Most often, the complete reaction energy accommodation is considered with the diffuse reflection of particles, which are products of the surface reaction~\cite{Molchanova-POF-2018}.  

\bigskip

\noindent{\bf   sTCE model for atomic species}

\smallskip
First the sTCE model is presented for a case most common in applications when the particle colliding with the surface is an atom.
In particular, in calculations of aerothermodynamics of the Earth atmosphere entry surface recombination of molecules is typically not considered, and therefore sTCE model for atomic species is mostly sufficient for such studies.

Let us consider a Cartesian coordinate system with the axis $X$ directed normal to a surface element. Then the equilibrium Maxwellian velocity distribution function of gas atoms of the species $A_g$ with the molecular mass $m$ near the surface can be written down as: 
\begin{equation}
f \left( \overrightarrow{V},T \right) = {\left( \frac{m}{2 \pi k T} \right)}^{\frac{3}{2}} \exp \left( -\frac{m \left( {V_n}^2 + {V_y}^2 + {V_z}^2 \right)}{2 k T} \right).
\label{eq:distr_func}
\end{equation}
where $\overrightarrow{V}=(V_n,V_y,V_z)$ is the atom velocity.
Let us assume that at the fixed constant surface density $n_{surf}$  the probability $P$ of an atom of the species $A_g$ colliding with a unit surface area to undergo a process similar to~\eqref{eq:adsorption} or~\eqref{eq:ERrecombination} is a function of the total collision energy $E$, that in the present variant of the model equals the translation energy of incoming atom $E=E_{tr}=mV^2/2$. Then the number of such processes per
unit time and surface area equals to:
\begin{equation}
n_{gas}\int\limits^{\infty}_0 \int\limits^{\infty}_0 \int\limits^{\infty}_0 P \left( E \right) \, V_n f \left( \overrightarrow{V}, T \right) \,dV_ndV_ydV_z.
\label{eq:number1}
\end{equation}

Equating the expression~\eqref{eq:number} and~\eqref{eq:number1} we obtain
\begin{equation}
K\left( T \right) n_{surf} = \int\limits^{\infty}_0 \int\limits^{\infty}_0 \int\limits^{\infty}_0 P \left( E \right) \, V_n f \left( \overrightarrow{V}, T \right) \,dV_n dV_y dV_z \, .
\label{eq:integral_equation_initial}
\end{equation}
Then the change of variables to spherical coordinates
$\{V_n, V_y, V_z\} \rightarrow \{V, \theta, \phi\}$ in the equation~\eqref{eq:integral_equation_initial} is performed:
\begin{equation}
V_n = V \mathrm{cos}\theta, \,\,\,\, \,\, 
V_y = V \mathrm{sin}\theta \mathrm{cos}\phi,  \,\, \,\,
V_z = V \mathrm{sin}\theta \mathrm{sin}\phi, 
\label{eq:change_variables}
\end{equation} 
where $\theta$ is the angle of incidence and $\phi$ is the azimuth angle. 
Next, by integrating over both angles and switching from integration over the speed $V$ to integration over the energy $E$, one obtains the following integral equation
\begin{eqnarray}
K\left( T \right) = &&\frac{1}{n_{surf}} {\left( \frac{1}{2 \pi m} \right)}^{1/2} {\left( \frac{1}{k T} \right)}^{3/2} \times \nonumber\\
&&\times \int\limits^{\infty}_0 P \left( E \right) \, E \exp \left( - \frac{E}{k T} \right) \,dE .
\label{eq:integral_equation}
\end{eqnarray}
Note that the integrand here is similar to the gas-phase TCE model (see Eq.~\eqref{eq:kTCE}) except for the second factor $E$ here (compare to $E^{\frac{\xi}{2}-1}$ in the gas-phase TCE), which reflects the fact that the equilibrium distribution function of translational energy of atoms coming to the surface is proportional to $E\exp(-E/kT)$ and hence has the form of the Boltzmann distribution for an energy mode with four degrees of freedom. 

Then similarly to the gas-phase TCE model we seek the probability in the following form
\begin{equation}
P \left( E \right) = C \dfrac{{\left( E - E_a \right)}^{\alpha}}{E} \mathcal{H} \left( E - E_a \right).
\label{eq:prob_micro_assumed}
\end{equation} 
Substituting~\eqref{eq:prob_micro_assumed} into the integral equation~\eqref{eq:integral_equation}, one obtains
\begin{equation} 
K \left( T\right) = \frac{C}{n_{surf}} \frac{1}{\sqrt{2 \pi m}} \mathrm{\Gamma} \left( \alpha + 1 \right) {\left( k T \right)}^{\alpha - \frac{1}{2}} \exp \left( -\frac{E_a}{k T} \right) .
\label{eq:solution}
\end{equation}
By equating the right-hand side of~\eqref{eq:solution} to the known $K(T)$ in the Arrhenius form~\eqref{eq:arrenius} one determines the sTCE probability of the impact surface process: 
\begin{equation} 
P \left( E \right) = \dfrac{A \sqrt{2 \pi m}}{\mathrm{\Gamma} \left( b + \frac{3}{2} \right) k^b} n_{surf} \dfrac{{\left( E - E_a \right)}^{b+\frac{1}{2}}}{E} \mathcal{H}\left( E - E_a \right).
\label{eq:prob_micro}
\end{equation} 

Note that this formula can be applied only when the integral on the right-hand side of Eq.~\eqref{eq:integral_equation} converges --- that is, when the condition $b > - \frac{3}{2}$ is satisfied.

\begin{table}[ht!]
\caption{Arrhenius parameters of surface processes}
\begin{center}
\label{table:rate_constants}
\begin{tabular}{ l | c | c | c}
Reaction & $A$ & $b$ & $E_a$ \\
\hline
$ \mathrm{O} + _S \longrightarrow \mathrm{O}_S $                    &  $5.01 \cdot 10^{-23}$ & 1.473 & $4.86 \cdot 10^{-22}$\\
$ \mathrm{O} + \mathrm{O}_S \longrightarrow \mathrm{O}_2 + _S $ 	&  $7.15 \cdot 10^{-23}$ & 1.484 & $2.78 \cdot 10^{-22}$ \\
\end{tabular}
\end{center}
\end{table}
\begin{figure}[!htpb]
\centering
\includegraphics[width=0.41\textwidth]{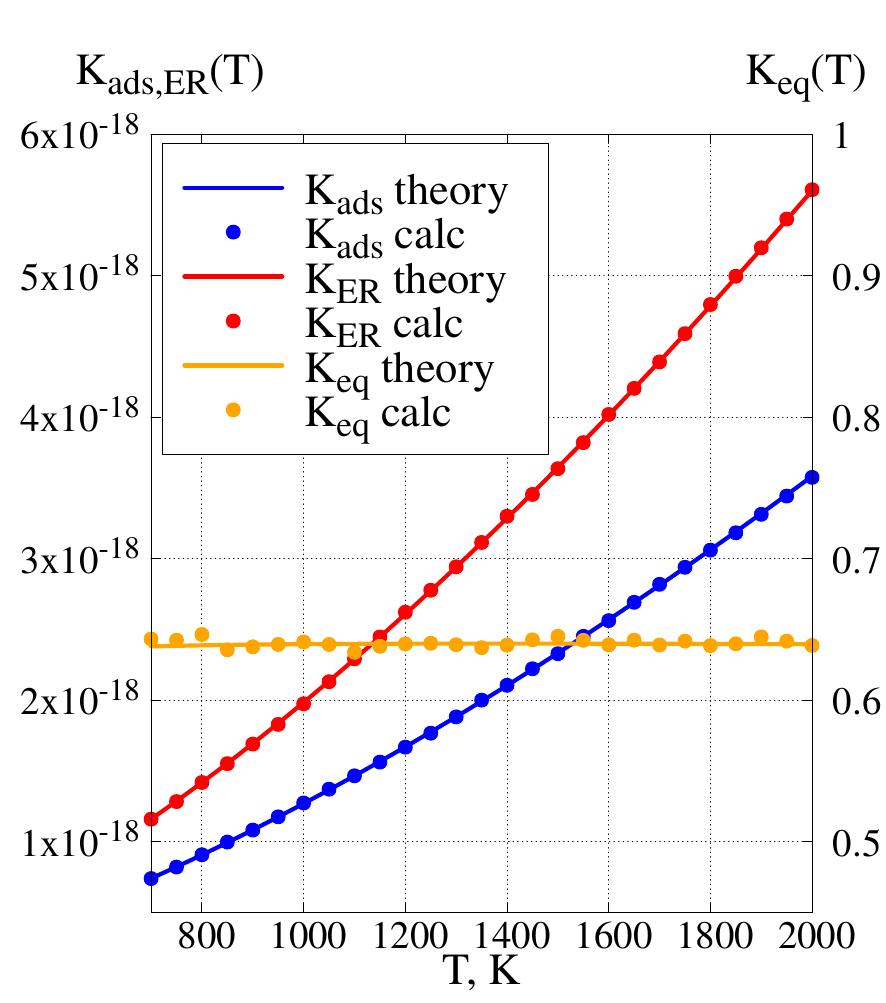}  
\caption[]{Comparison of the DSMC results (symbols) with  $K_{ads}$, $K_{ER}$ and $K_{eq}$ (curves).
        }
\label{fig:verification}
\end{figure}

Figure~\ref{fig:verification} presents results of verification of implementation of the sTCE model with
the  probability formula~\eqref{eq:prob_micro} into the SMILE++ DSMC software system~\cite{SMILE++RGD27} of the SMILE family~\cite{SMILE-family}.
The processes of oxygen atom adsorption and Eley-Rideal recombination on the $\beta$-cristobalite surface were selected for the numerical tests with the rate constant data~\cite{Kroupnov-ActaAstronautica-2023} given in Table~\ref{table:rate_constants}.

In the first series of DSMC tests the number of reactions was counted, but the reactions themselves were not actually modeled, so the surface coverage did not change.
Numerical results for this series for the adsorption and Eley-Rideal recombination rate constants shown in blue and red symbols, respectively, are compared with the theoretical values (curves).
In the second series of tests the reactions were modeled and the surface coverage was not fixed. The  numerical results of this series for the equilibrium constant $K_{eq} = K_{ads}/K_{ER}$ and its theoretical values are also shown in Fig.~\ref{fig:verification}.
Excellent agreement is observed between the theoretical values and numerical results, confirming the correct implementation of the sTCE model in the SMILE++ system.

\bigskip

\noindent{\bf   Comparison of probability expressions for sTCE model and model~\cite{Molchanova-POF-2018}}

\begin{figure*}[!htpb]
\centering
{\bf sTCE probability function, Eq.~\eqref{eq:prob_micro}}

\subfloat[] {\includegraphics[width=0.19\textwidth]{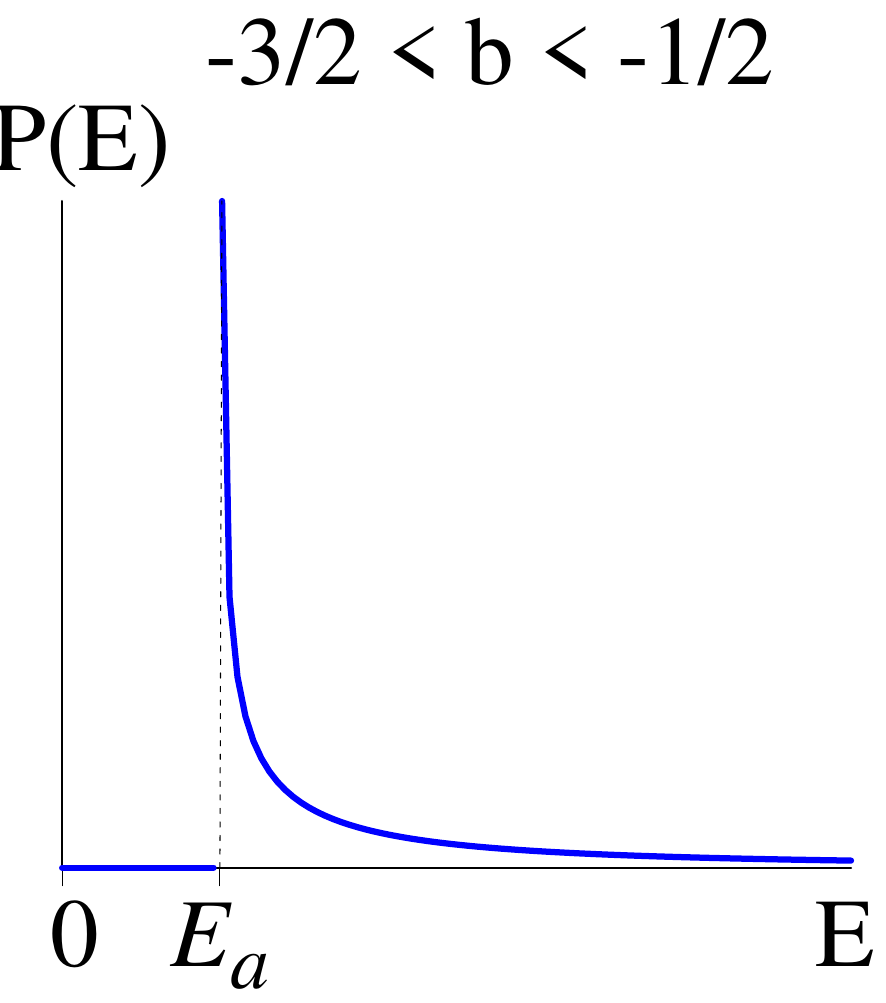}  \label{fig:prob_E_1}}
\subfloat[] {\includegraphics[width=0.19\textwidth]{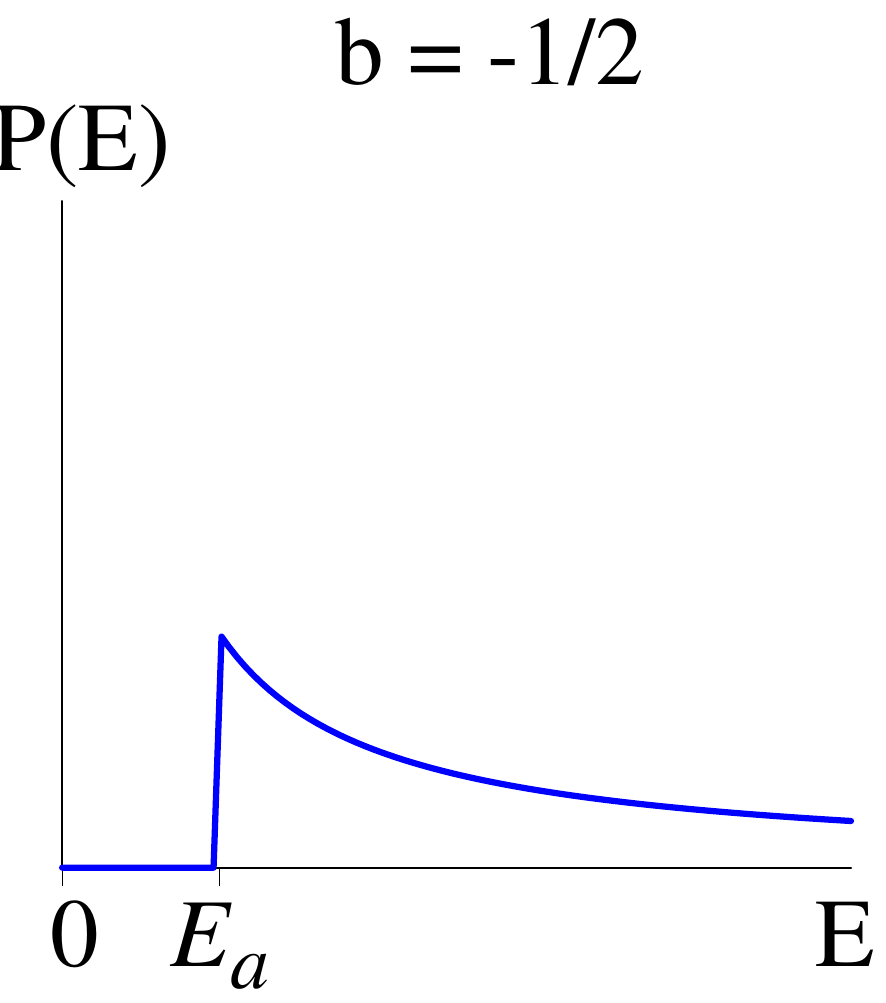}  \label{fig:prob_E_2}}
\subfloat[] {\includegraphics[width=0.19\textwidth]{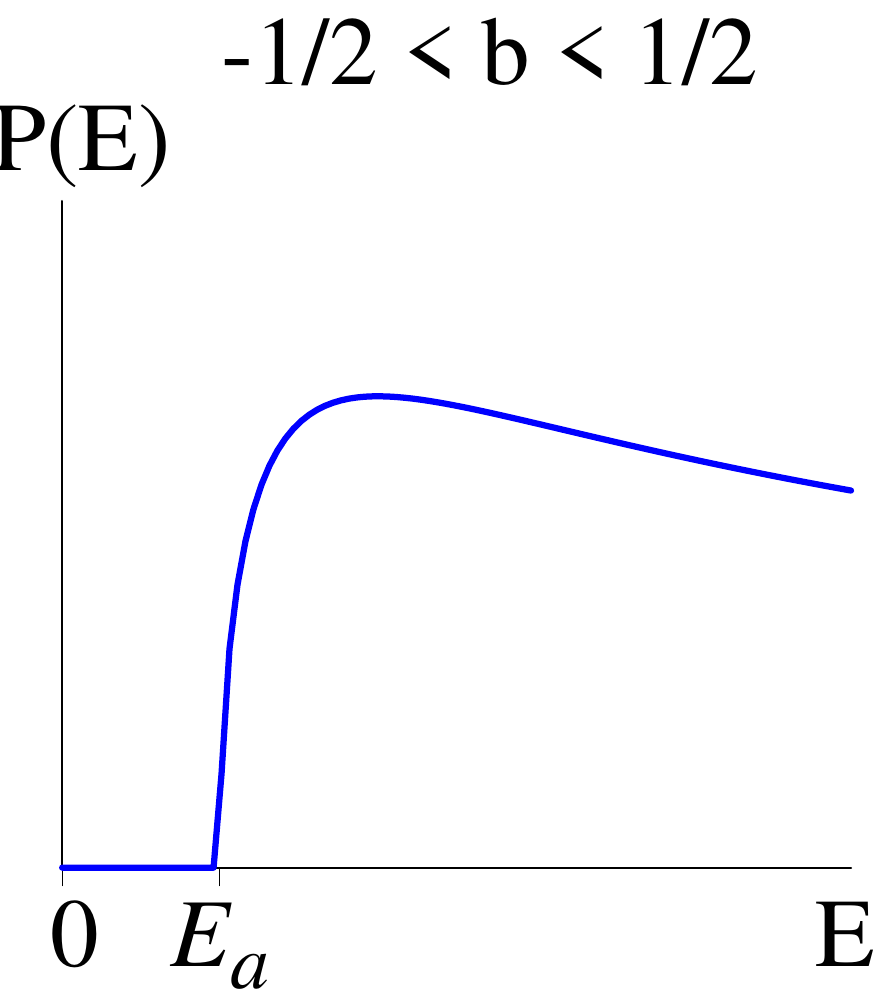}  \label{fig:prob_E_3}}
\subfloat[] {\includegraphics[width=0.19\textwidth]{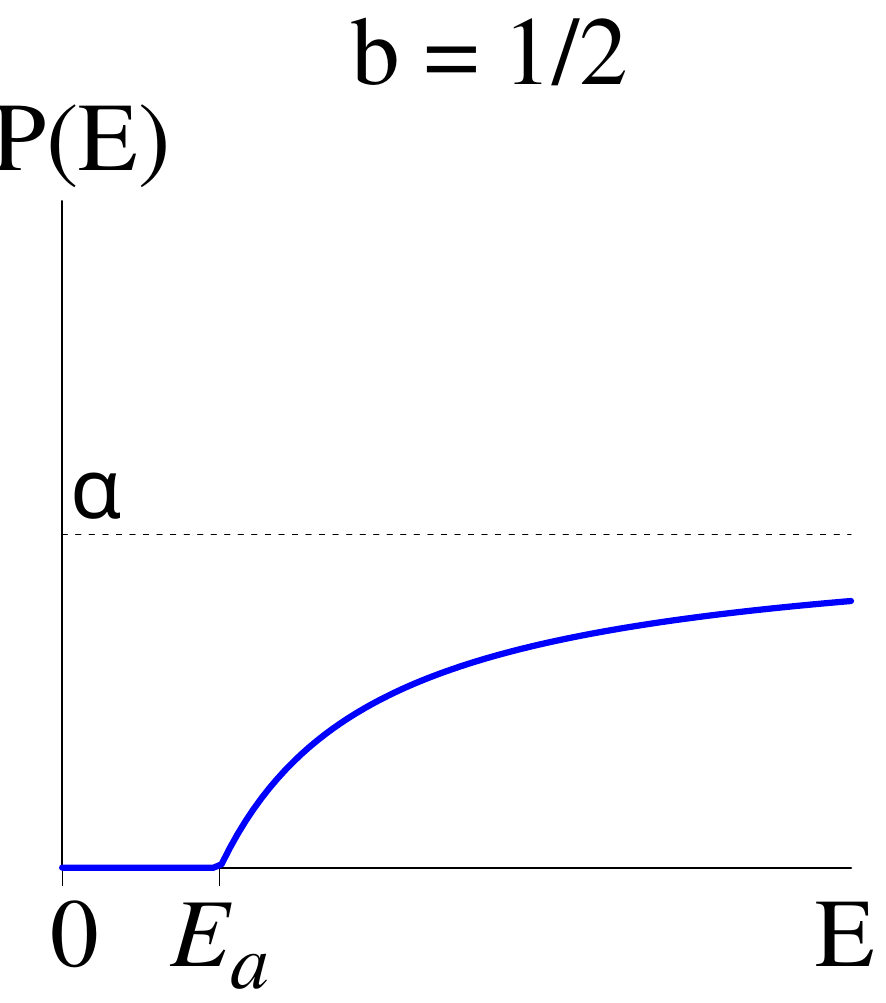}  \label{fig:prob_E_4}}
\subfloat[] {\includegraphics[width=0.19\textwidth]{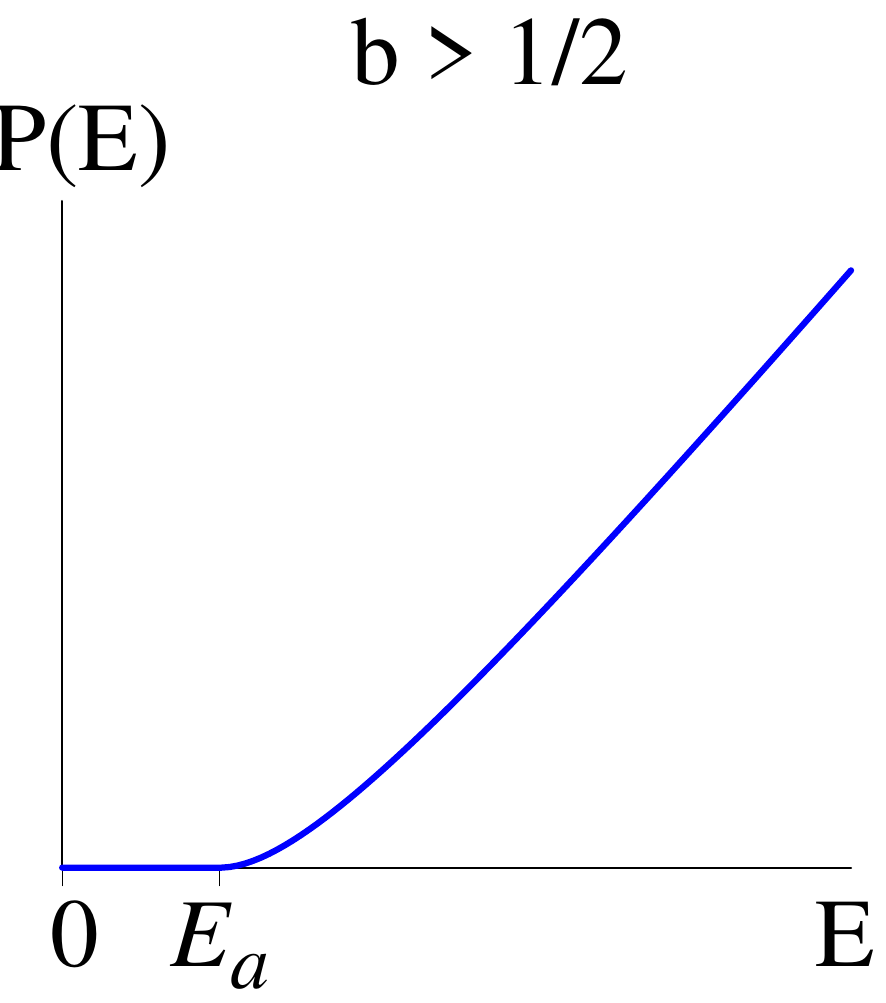}  \label{fig:prob_E_5}}

{\bf Probability function of model~\cite{Molchanova-POF-2018}, Eq.~\eqref{eq:prob_micro_En}}

\subfloat[] {\includegraphics[width=0.19\textwidth]{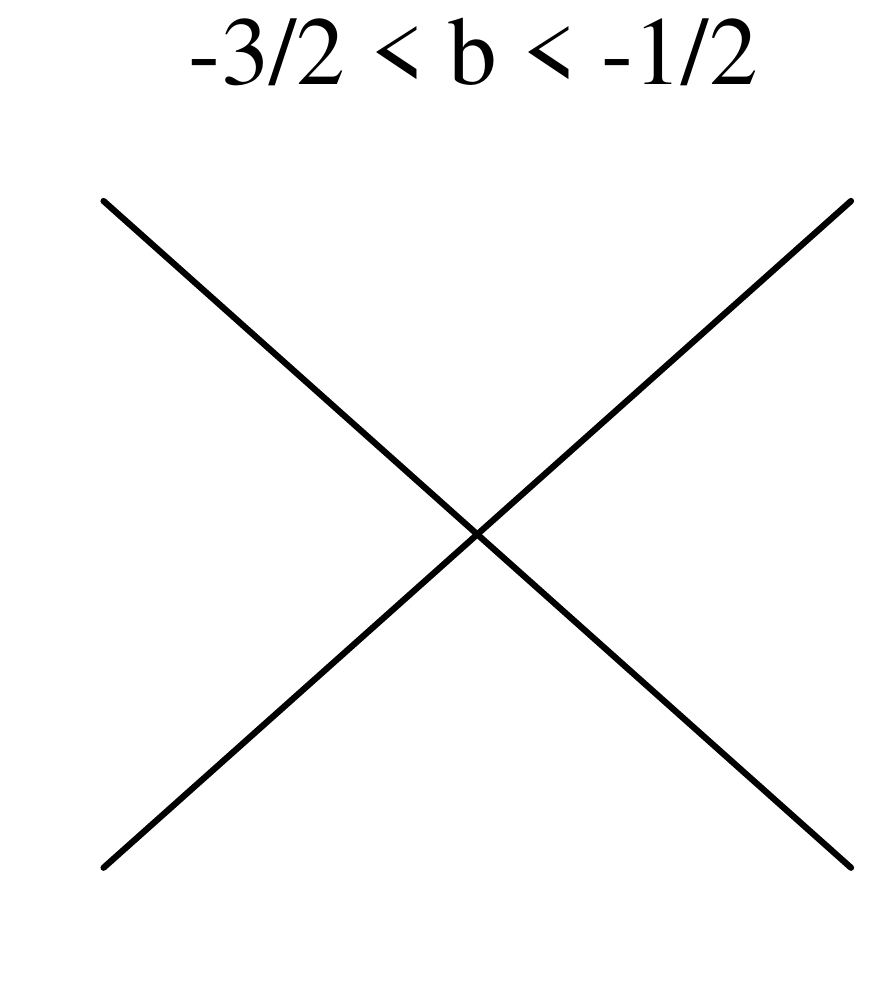}  \label{fig:prob_En_1}}
\subfloat[] {\includegraphics[width=0.19\textwidth]{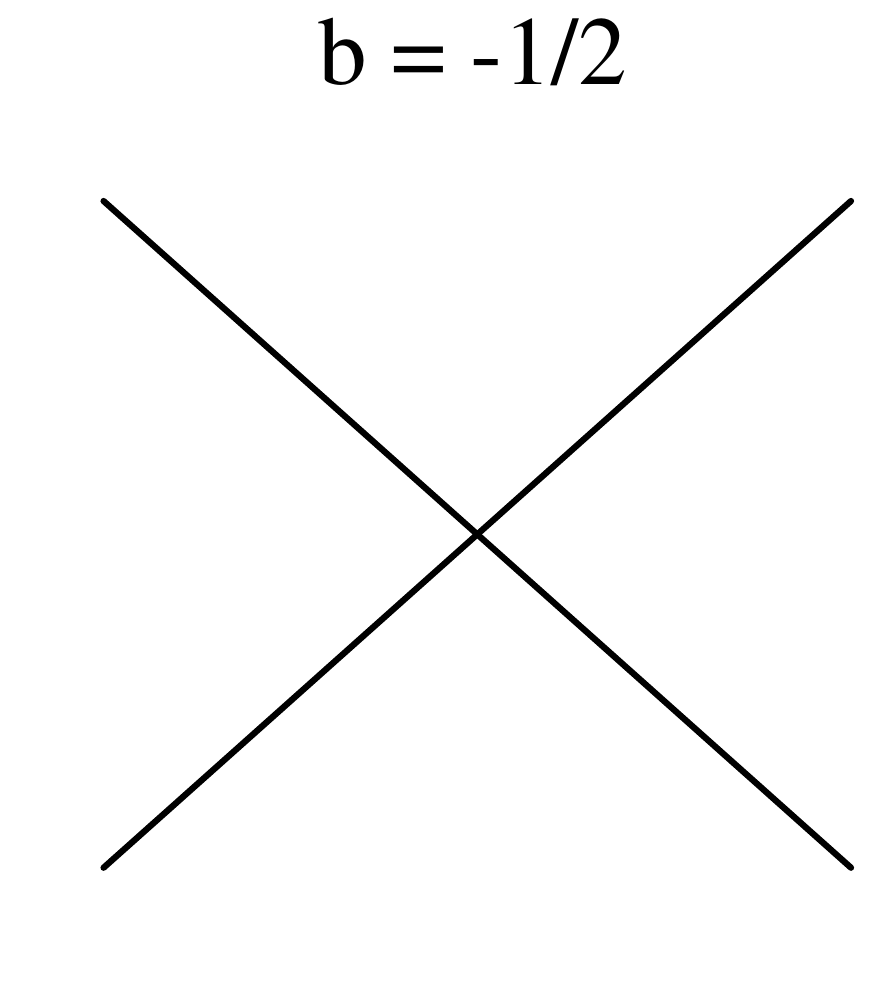}  \label{fig:prob_En_2}}
\subfloat[] {\includegraphics[width=0.19\textwidth]{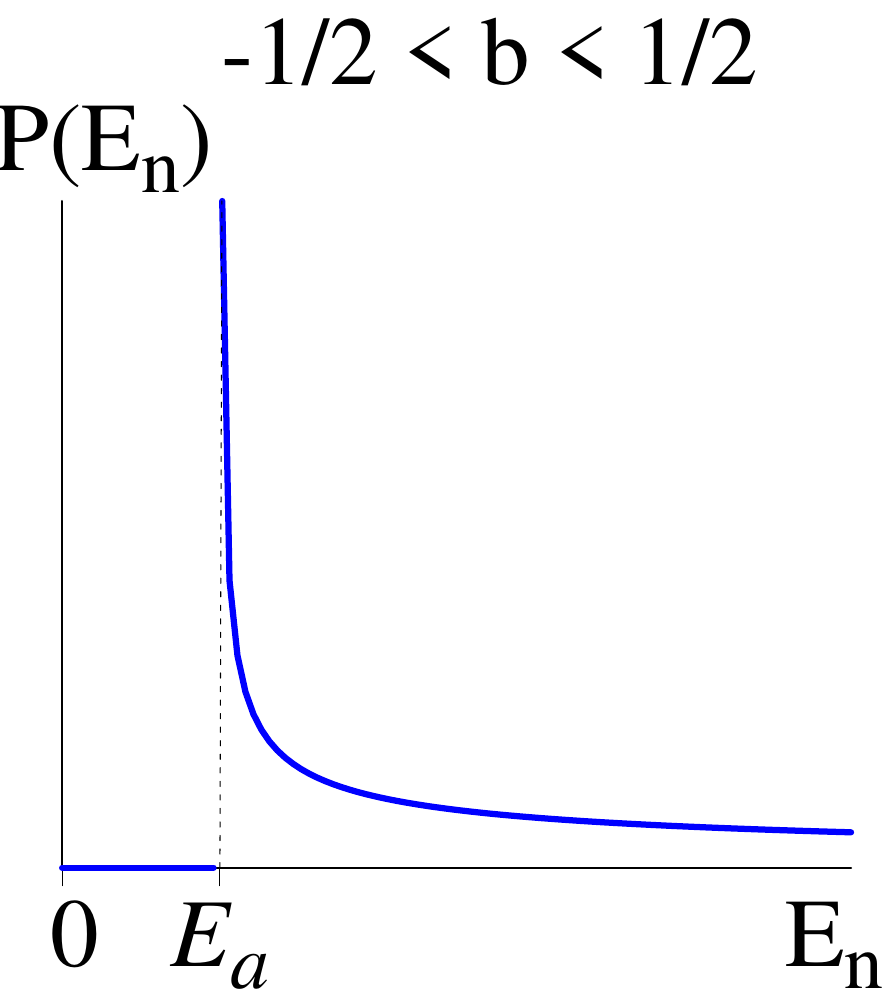}  \label{fig:prob_En_3}}
\subfloat[] {\includegraphics[width=0.19\textwidth]{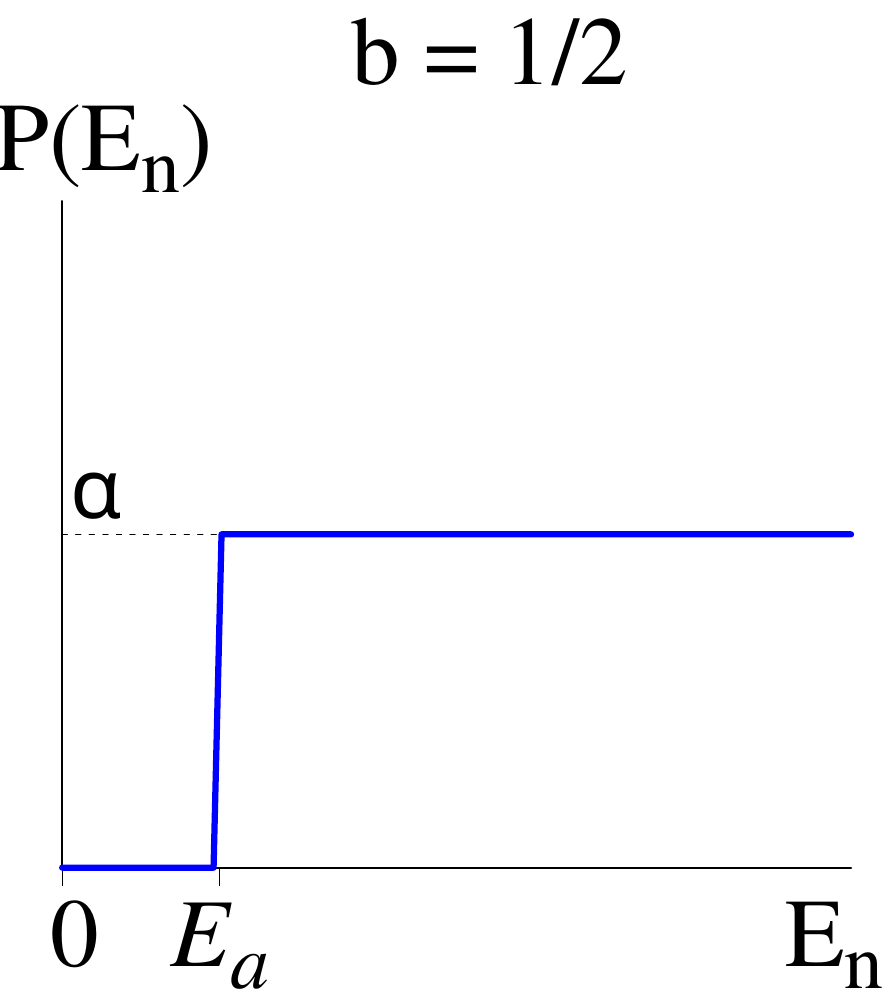}  \label{fig:prob_En_4}}
\subfloat[] {\includegraphics[width=0.19\textwidth]{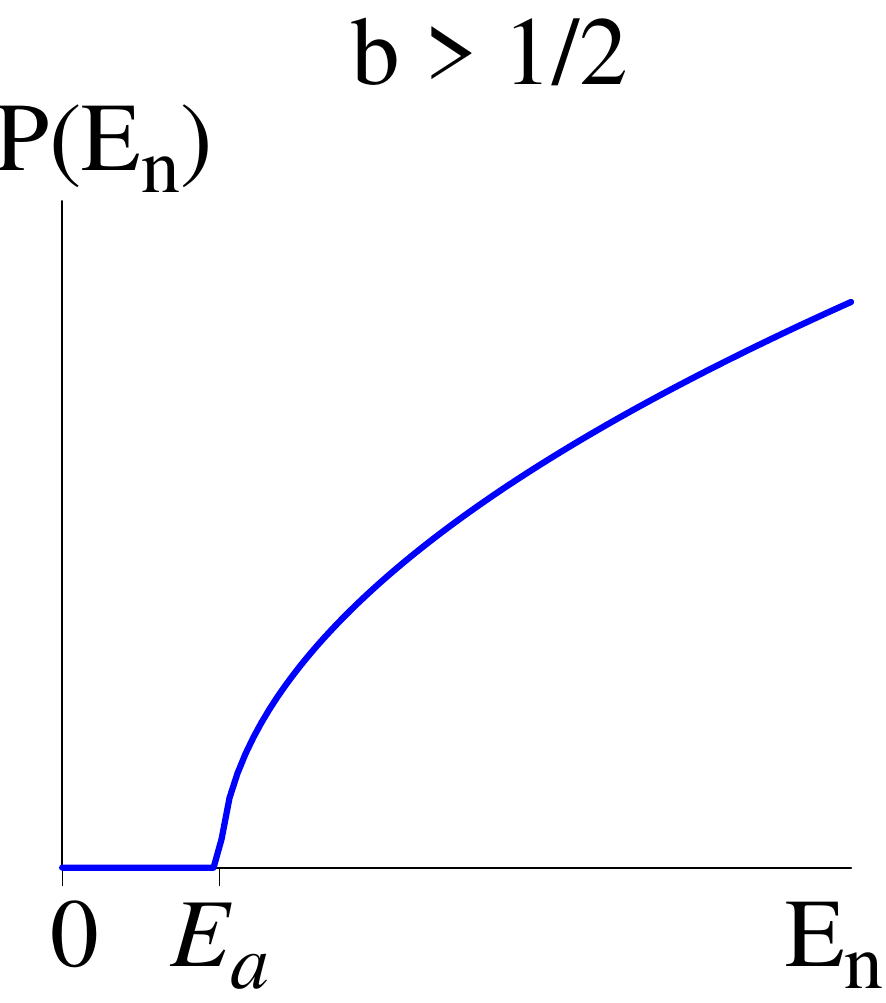}  \label{fig:prob_En_5}}

\caption[]{Forms of probability functions $ P \left( E \right) $ in the presented sTCE model and $ P \left( E_{n} \right) $ in the model~\cite{Molchanova-POF-2018} for various $b$.
        }
\label{fig:prob_micro}
\end{figure*}

\smallskip
\noindent
In the study~\cite{Molchanova-POF-2018}, it was assumed that the probability of the surface process depends on the normal component of the velocity of a particle impinging on the surface similarly to the previous work~\cite{Choquet-1999}.
If the expression for the probability~\cite{Molchanova-POF-2018} is reformulated in terms of the normal component of translational energy $ E_{n} = mV_n^2/2 $, it takes the form:
\begin{equation} 
P \left( E_n \right) = \dfrac{A \sqrt{2 \pi m}}{\mathrm{\Gamma} \left( b + \frac{1}{2} \right) k^b} n_{surf} {\left( E_n - E_a \right)}^{b-\frac{1}{2}} \mathcal{H} \left( E_n - E_a \right).
\label{eq:prob_micro_En}
\end{equation}

Note that this expression correspond to the TCE model with $\xi=2$. Indeed, if one considers $E_n$ the total energy which contributes to the process probability and takes into account that the equilibrium distribution of $E_n$ is proportional to $\exp(-E_n/kT)$ (and therefore is the Boltzmann distribution for two degrees of freedom), then one obtains the TCE probability which exactly matches Eq.~\eqref{eq:prob_micro_En}.
The condition for the convergence of the integral implies that this formula is applicable provided that $b > - \frac{1}{2}$.

A comparison of the typical energy dependence of the probability for various values of the parameter $b$ for the sTCE model (Eq.~\eqref{eq:prob_micro}) and model~\cite{Molchanova-POF-2018} (Eq.~\eqref{eq:prob_micro_En}) is presented in Fig.~\ref{fig:prob_micro}.
As noted above, the sTCE model is applicable for $b > -3/2$, whereas the model~\cite{Molchanova-POF-2018} can be used for $b > -1/2$.
The energy dependence of the probability for the sTCE model with $b$ in the interval $(-\frac{3}{2}, -\frac{1}{2})$, as well as for the model~\cite{Molchanova-POF-2018} with $b$ in the interval $(-\frac{1}{2}, \frac{1}{2})$, is unbounded in the vicinity of the reaction threshold $E_a$.
Since the function $P$ represents a probability, the condition $P < 1$ must be satisfied for DSMC computations to yield correct equilibrium reaction rates.
This issue has been discussed both in the context of the gas-phase~\cite{Accuracy-RGD24-2005} and surface~\cite{Molchanova-Lobachevskiy-2026} chemistry.
The calculations of the reaction rate using the DSMC method for the sTCE model actually correspond in thermal equilibrium to the integral~\eqref{eq:integral_equation} with the probability $P$ truncated by the value of 1.
It should be noted that the study~\cite{Molchanova-POF-2018} examined reactions with parameter $b$ values in the range $( -\frac{1}{2}, \frac{1}{2})$ and a significant discrepancy between the DSMC results for the model~\cite{Molchanova-POF-2018} and the equilibrium Arrhenius rate was obtained.
This problem was resolved only by using a function $P$ that depends on both energy and surface temperature. This approach will be discussed further.

It should be noted that for both models, the probability function $P$ is increasing for $b > 1/2$ and, consequently, also exceeds unity at high energies.
However, since the equilibrium energy distribution function decreases exponentially at infinity, the contribution of high-energy molecules to the calculation of integral~\eqref{eq:integral_equation} for the sTCE model or the similar integral for model~\cite{Molchanova-POF-2018} is small. Consequently, the use of these models at such values of $b$ does not lead to a significant error.
This statement is supported by the  successful verification results for the sTCE model with $b$ values close to 1.5 presented above. 

Thus, it can be asserted that the use of models is fully valid only when the probability function is bounded in the vicinity of the reaction threshold. That is the sTCE model is valid for $b \ge -1/2$, and the model~\cite{Molchanova-POF-2018} is valid for $b \ge 1/2$.
The range of valid application of the sTCE model is broader and encompasses the particularly important case of $b=0$, since data on rate constants with $b=0$ are widely prevalent in the literature.

\bigskip

\noindent{\bf   sTCE probabilities dependent on energy and surface temperature}

\smallskip
\noindent
The probability of the surface process may depend not only on the energy of the incoming particle but on the surface temperature as well~\cite{SurfTemp,Scaling}. The extension of the sTCE model which allows for the dependence of the probability on the energy and wall temperature is presented below.

Let us assume that the dependence of the probability on the surface temperature has the Arrhenius form, so the probability function can be presented as follows: 
\begin{equation}  
P_{E,T}(E,T_w) = P(E) A_w T_w^{b_w} \exp \left( -\frac{E_w}{k T_w} \right)  ,
\label{ProbET}
\end{equation}
where $T_w$ is the surface temperature and $A_w$, $b_w$ and $E_w$ are constants. If one sets the values of these constants, then in thermal equilibrium, when gas temperature equals surface temperature $T=T_w$, Eq.~\eqref{eq:integral_equation} is valid with $K(T)$ substituted with the function
\begin{equation}
\widetilde{K}(T)=\widetilde{A} {T}^{\widetilde{b}} \exp \left( -\frac{\widetilde{E_a}}{k T} \right), 
\end{equation}
where $\widetilde{A} = A/A_w$, $\widetilde{b} = b - b_w$, and $\widetilde{E_a} = E_a - E_w$. By solving the equation one determines $P(E)$ as
\begin{equation}
P \left( E \right) = \dfrac{\widetilde{A} \sqrt{2 \pi m}}{\mathrm{\Gamma} \left( \widetilde{b} + \frac{3}{2} \right) k^{\widetilde{b}}} n_{surf} \dfrac{{\left( E - \widetilde{E_a} \right)}^{\widetilde{b}+\frac{1}{2}}}{E} \mathcal{H} \left( E - \widetilde{E_a} \right).
\label{eq:prob_microET}
\end{equation}

Clearly the probability $P_{E,T}(E,T_w)$ does not depend on $A_w$, which can therefore be set to unity without loss of generality. In this case $\widetilde{A}=A$. The choice of other two parameters $b_w$ and $E_w$ cannot be done only on the basis of the macroscopic reaction rates and requires some insights from theory, computer chemistry or experiments.  Additionally there are mathematical limitations on the values of these parameters. Taking into account the considerations made in the previous sections it can be seen that probability function~\eqref{eq:prob_microET} is valid when
\begin{equation} 
\widetilde{b} = b - b_w> - \frac{3}{2},
\end{equation}
and is limited in the vicinity of the reaction threshold when
\begin{equation} 
\widetilde{b} = b - b_w \ge - \frac{1}{2}.
\label{1234}
\end{equation}
The limitation on the parameter  $E_w$ follows from the non-negativity of $\widetilde{E_a}$ required for the integral equation solution $\widetilde{E_a} = E_a - E_w \ge 0.$ Note that the form of $P(E)$ given by Eq.~\eqref{eq:prob_microET} can be presented by Figures~\ref{fig:prob_E_1}-e if $E_a$ is substituted with  $\widetilde{E_a}$ and $b$ with  $\widetilde{b}$.

Both parameters $b_w$ and $E_w$ preferably should be chosen on the basis of available data. However, pure mathematical considerations may be also applied. For example, if $b < -1/2$, then the original sTCE model ($b_w = E_w = 0$) cannot be used at all or probability function given by  Eq.~\eqref{eq:prob_micro}  has the problematic form discussed earlier as one shown in Fig.~\ref{fig:prob_E_1}. In this case one can choose a negative $b_w$ in order to fulfill the condition given by Eq.~\eqref{1234} and overcome this mathematical issue. In fact, in paper~\cite{Molchanova-POF-2018} due to numerical problems with probability function of the form shown in Fig.~\ref{fig:prob_En_3}, the variant of $P_{E,T}(E_n,T_w)$ probability was employed with $E_w=E_a$ which helped to overcome the problems by shifting the reaction threshold to zero.

\bigskip

\noindent{\bf   sTCE probabilities for specified reaction efficiency models}

\smallskip
\noindent
In CFD studies of spacecraft aerothermodynamics relatively simple specified reaction efficiency (SRE) models~\cite{Marschall-ChapterInBook-2015} instead of detailed finite-rate surface reaction models are typically used for surface catalytic reactions.
In these models, the entire set of physical and chemical elementary steps is described by mutually independent irreversible reactions of the form:
\begin{equation}
\mathrm{A} + \mathrm{A} \longrightarrow \mathrm{A_{2}},
\end{equation}
where $\mathrm{A}$ typically represents nitrogen or oxygen atoms impinging on the surface, and A$_2$ represents nitrogen or oxygen molecules emitted from the surface into the gas phase.

Such models are easily implemented within the framework of the DSMC method~\cite{Shumakova-AIAA-2014}.
The following algorithm is used for the interaction between an atom and a surface element.
If there is no adsorbed atom of the same species on the surface, the atom is adsorbed by a surface element with probability $P$.
If an adsorbed atom is present on the surface, the atoms recombine into a molecule with the same probability $P$, and the molecule is emitted into the gas phase.
The released recombination energy is distributed between the surface and the energy modes of the molecule in accordance with a prescribed law.
It should be noted that constant or depending only on the surface temperature values of the probability $P$ are typically used.
The  sTCE approach allows one to obtain probabilities depending on the energy of the incident atom.

At the macroscopic level, within the framework of SRE models, the surface reaction rate is determined either by the recombination rate constant $K_w(T)$ [m/s] or by the dimensionless recombination coefficient (catalytic efficiency) $\gamma(T)$, representing the average probability of recombination for an atom impinging the surface.
The case of $\gamma=0$ corresponds to a non-catalytic wall, while $\gamma=1$ corresponds to a fully catalytic wall.
Under the thermal equilibrium conditions the number of recombination reactions of atoms $A$ per unit surface area per unit time, expressed in terms of these parameters, is equal to
\begin{equation}
K_w(T)n_{gas} =  \gamma \left( T \right)   \sqrt{\frac{k T}{2 \pi m}} n_{gas},
\label{111}
\end{equation}
where $n_{gas}$ and $m$ are the gas-phase density and the mass of atoms A.

Applying the sTCE approach, one obtains the following analogues of equation~\eqref{eq:integral_equation}:
\begin{equation}
\gamma\left( T \right) =  {\left( \frac{1}{k T} \right)}^{{2}} 
\int\limits^{\infty}_0 P \left( E \right) \, E \exp \left( - \frac{E}{k T} \right) \,dE,
\label{eq:integral_equation1}
\end{equation}
\begin{equation}
K_w\left( T \right) = {\left( \frac{1}{2 \pi m} \right)}^{\frac{1}{2}} {\left( \frac{1}{k T} \right)}^{\frac{3}{2}} 
\int\limits^{\infty}_0 P \left( E \right) \, E \exp \left( - \frac{E}{k T} \right) \,dE.
\label{eq:integral_equation2}
\end{equation}

If the recombination coefficient $\gamma(T)$ is given in the Arrhenius form~\eqref{eq:arrenius},then the solution of Eq.~\eqref{eq:integral_equation1} is
\begin{equation}
P \left( E \right) = \dfrac{A}{\mathrm{\Gamma} \left( b + 2 \right) k^{b}} \dfrac{{\left( E - E_{a} \right)}^{b + 1}}{E} \mathcal{H} \left( E - E_{a} \right).
\label{eq:prob_micro_gamma}
\end{equation}
This formula is applicable for $b > -2$ with the probability bounded in the vicinity of the reaction threshold for $b \geq -1$.

If the recombination rate constant $K_w \left( T \right)$ is given in the Arrhenius form~\eqref{eq:arrenius}, then one obtains the solution of Eq.~\eqref{eq:integral_equation2}:
\begin{equation}
P \left( E \right) = \dfrac{A \sqrt{2 \pi m}}{\mathrm{\Gamma} \left( b + \frac{3}{2} \right) k^{b}} \dfrac{{\left( E - E_a \right)}^{b+\frac{1}{2}}}{E} \mathcal{H} \left( E - E_a \right).
\label{eq:prob_micro_Kw}
\end{equation}
The conditions for the applicability of this formula are the same as for Eq.~\eqref{eq:prob_micro}.

Note that the relationship between $K_w$ and $\gamma$ in Eq.~\eqref{111} follows from the Maxwellian form of the velocity distribution function near the surface.
Some macroscopic models assume a non-equilibrium form of the distribution function, which leads to different relations connecting these quantities (however in some cases Eq.~\eqref{111} remains valid in thermal nonequilibrium)~\cite{Scott}.
For this reason, the sTCE approach for the SRE formulation should be used with caution: particularly, it is necessary to ensure that the probability expressions are not based on contradictory assumptions regarding the velocity distribution function.

\bigskip

\noindent{\bf   sTCE probabilities with dependence on the incidence angle}

\smallskip
\noindent
Probability of impact surface process may significantly depend~\cite{Bonfanti-QuantumChem-2016} on the angle of incidence of the incoming atom $\theta$. 
Then the probability $P_{E,\theta} \left(E ,  \theta \right)$ can be found as a solution of the following integral equation: 
\begin{eqnarray}
\label{eq:integral_equation_ang}
K\left( T \right) = \frac{1}{n_{surf}} {\left( \frac{2}{\pi m} \right)}^{1/2} {\left( \frac{1}{k T} \right)}^{3/2} \times \\
\times \int\limits^{\pi/2}_0 \int\limits^{\infty}_0 P_{E,\theta} \left( E ,  \theta \right) \, E \exp \left( - \frac{E}{k T} \right) \,dE \, \sin \theta \cos \theta \,d\theta. \nonumber
\end{eqnarray}

One of the possible solutions of this equation has the following form with arbitrary parameter $\lambda>-2$: 
\begin{equation}
P_{E,\theta} \left( E ,  \theta \right) = P \left( E \right) \, \frac{\lambda+2}{2}  \cos^{\lambda} \theta  ,
\label{eq:prob_from_angle}
\end{equation}
where $P(E)$ is given by Eq.~\eqref{eq:prob_micro}. Negative values of $\lambda$ should be avoided due to unlimited probability for high angles close to $\pi/2$. 

When angular dependence of reaction probability is discussed, often two limiting cases are considered: total energy scaling (TES) when the probability depends on total translational energy $E$,  and normal energy scaling (NES) when the probability is a function of the normal translational energy $E_n=E \cos^2 \theta$. It has been suggested~\cite{Scaling, Bonfanti-QuantumChem-2016} that many processes which exhibit neither TES nor NES behavior scale on $E_{\chi}=E \cos^{\chi} \theta$, where $\chi$ is a scaling parameter:  $0 \leq \chi \leq 2$. For the probability function of a form: 
\begin{equation}
P_{E,\theta} \left( E ,  \theta \right) = P(E \cos^{\chi} \theta) = P(E_{\chi}) 
\label{eq:scaling}
\end{equation}
another solution of Eq.~\eqref{eq:integral_equation_ang} can be obtained (see {\it Supplementary materials} for the strict derivation):  
\begin{eqnarray}
P\left(E_{\chi}\right) =
\dfrac{ A \sqrt{\pi m/2}}{\mathrm{\Gamma} \left( b + \frac{3}{2} \right) k^b} n_{surf} \dfrac{{\left( E_{\chi} - E_a \right)}^{b-\frac{1}{2}}}{E_{\chi}} \times \nonumber \\
\times \chi \bigg[(b-\frac{1}{2}+\frac{2}{\chi})E_{\chi}+ (1-\frac{2}{\chi})E_a   \bigg] \mathcal{H} (E_{\chi}-E_a). 
\label{eq:prob_from_angle1}
\end{eqnarray}
This expression coincides in the NES limit ($\chi=2$) with Eq.~\eqref{eq:prob_micro_En} and in the TES limit ($\chi \rightarrow 0$) with the original sTCE probability \eqref{eq:prob_micro}. However for all values of $\chi$ in $(0,2]$ it has the same limitations in terms of $b$ as probability~\eqref{eq:prob_micro_En} which are more strict than for the original sTCE probability and lead to exclusion 
of the important case of $b=0$.

\bigskip

\noindent{\bf   sTCE model for molecular species}

\smallskip
\noindent
The sTCE model can be generalized to include the contribution of the internal energy of molecules to the reaction probability. Let us consider an impact surface process involving a molecule AB$_g$ impinging on a surface (e.g., adsorption, Eley-Rideal recombination, dissociation, etc.) 

It is assumed that the probability of the process $P$ depends on the total energy $E$, which is equal to the sum of the translational energy $E_{tr}$ and the internal (rotational-vibrational) energy $E_i$ of the molecule: $E = E_{tr} + E_i$.
For this case, the analogue of Eq.~\eqref{eq:integral_equation} takes the following form
\begin{eqnarray}
&&K\left(  T \right) = \frac{1}{n_{surf}} {\left( \frac{1}{2 \pi m} \right)}^{1/2} {\left( \frac{1}{k T} \right)}^{3/2} \times \nonumber\\
&&\times \int\limits^{\infty}_0 \int\limits^{\infty}_0 P \left( E \right) \, E_{tr} \exp \left( - \frac{E_{tr}}{k T} \right) f \left( E_i,T \right) \, dE_{tr} dE_i \, ,
\label{eq:integral_equation_molecule}
\end{eqnarray}
where the internal energy distribution function is written as
\begin{equation}
f \left( E_i,T \right) = \frac{1}{\mathrm{\Gamma} \left( \xi_i/{2} \right) k T} {\left( \frac{E_i}{k T} \right)}^{\frac{\xi_i}{2}-1} \exp \left( -\frac{E_i}{k T} \right) \, .
\label{eq:distr_func_molecule}
\end{equation}
Here,  $\xi_i$ is the number of internal degrees of freedom of the molecule. As in the original TCE model, rotational and vibrational energies are assumed to be continuous.
Similarly to Eq. \eqref{eq:prob_micro}, we obtain the probability of the process
\begin{eqnarray}
P \left( E \right) = &&
A  \dfrac{ \Gamma \left(2+ \frac{{\xi}_i}{2} \right) }{\mathrm{\Gamma} \left( b + \frac{3}{2} + \frac{{\xi}_i}{2} \right)} \dfrac{\sqrt{2 \pi m}}{k^b} 
\times \nonumber \\
&&\times n_{surf} \dfrac{{\left( E - E_a \right)}^{b+\frac{1}{2}+\frac{{\xi}_i}{2}}}{{E^{ \frac{{\xi}_i}{2} + 1}}} \mathcal{H} \left( E - E_a \right).
\label{eq:prob_micro_molecule}
\end{eqnarray}
This formula is applicable when $b + {\xi}_i/2 > 3/2$ and does not yield an infinite probability in the vicinity of the reaction threshold when $b + {\xi}_i/2 \geq 1/2$.

It should be noted that the sTCE model for molecules has the same limitations and features as the TCE model for gas-phase reactions~\cite{Gimelshein-BirdLegacy-2019}.
In particular, the temperature dependence of ${\xi}_i$ leads to a temperature dependence of the probability~\eqref{eq:prob_micro_molecule}.
During DSMC computations, the temperature can be calculated by sampling the parameters of incident and reflected particles or assumed to be equal to the wall temperature.
Another limitation relates to the use of this probability formulae for discrete rotational and vibrational energies.
As with the original TCE model, this may require adjusting the parameters~\cite{Gimelshein-PhysFluids-2004} appearing in formula~\eqref{eq:prob_micro_molecule}.

\bigskip

\noindent{\bf   Summary and conclusions}

\smallskip
\noindent
The total collision energy (TCE) approach which is conventional for the gas phase chemistry within the DSMC community has been applied to surface chemical reactions. 
The proposed sTCE model can be applied within a DSMC code to an elementary-step process involving an atom impinging on a surface with a reaction rate given by the modified Arrhenius equation. The probability of the process given in the analytical form depends on the the translational energy of the atom.
The sTCE can be used for much higher range of the Arrhenius temperature exponent parameter $b$ than previous models making it better suited for  applications. Implementation of the model in the SMILE++ DSMC software system demonstrates total agreement of the calculation results with the theory under thermal equilibrium conditions.

Several generalizations of the sTCE model have been proposed, namely, for taking into account the effect of surface temperature, incidence angle, and rotational and vibrational energy for processes involving molecules incident on the surface. A variant of the model has been suggested for using it within the specified reaction efficiency approach, which is more feasible in the engineering context than more complicated finite-rate formulation. Most of these generalizations of the sTCE model can be used in combination with each other, e.g. it is possible to implement  
 a sTCE probability function, which simultaneously depend on  translational energy of an incident atom, angle of its incidence and surface temperature, in a DSMC code. Insights from computational chemistry and experimental data could be crucial for obtaining parameters of such combined sTCE models with multiple dependencies.      

\smallskip
\noindent{\bf Acknowledgments}. The sTCE model development 
was supported by the Russian Science Foundation (Grant No.
25-79-30031). Software implementation of the model was conducted within the State Assignment of ITAM SB RAS (Project No. 124021400040-4). 

\smallskip
\noindent{\bf Supplementary materials} present derivation of Eq.~\eqref{eq:prob_from_angle1}.

\smallskip
\noindent{\bf Data availability}. The data that support the findings of this study are available from
the corresponding author upon reasonable request.
\smallskip

\bigskip
\noindent{\bf REFERENCES}
\smallskip


\begin{thebibliography}{23}%
\makeatletter
\providecommand \@ifxundefined [1]{%
 \@ifx{#1\undefined}
}%
\providecommand \@ifnum [1]{%
 \ifnum #1\expandafter \@firstoftwo
 \else \expandafter \@secondoftwo
 \fi
}%
\providecommand \@ifx [1]{%
 \ifx #1\expandafter \@firstoftwo
 \else \expandafter \@secondoftwo
 \fi
}%
\providecommand \natexlab [1]{#1}%
\providecommand \enquote  [1]{``#1''}%
\providecommand \bibnamefont  [1]{#1}%
\providecommand \bibfnamefont [1]{#1}%
\providecommand \citenamefont [1]{#1}%
\providecommand \href@noop [0]{\@secondoftwo}%
\providecommand \href [0]{\begingroup \@sanitize@url \@href}%
\providecommand \@href[1]{\@@startlink{#1}\@@href}%
\providecommand \@@href[1]{\endgroup#1\@@endlink}%
\providecommand \@sanitize@url [0]{\catcode `\\12\catcode `\$12\catcode
  `\&12\catcode `\#12\catcode `\^12\catcode `\_12\catcode `\%12\relax}%
\providecommand \@@startlink[1]{}%
\providecommand \@@endlink[0]{}%
\providecommand \url  [0]{\begingroup\@sanitize@url \@url }%
\providecommand \@url [1]{\endgroup\@href {#1}{\urlprefix }}%
\providecommand \urlprefix  [0]{URL }%
\providecommand \Eprint [0]{\href }%
\providecommand \doibase [0]{http://dx.doi.org/}%
\providecommand \selectlanguage [0]{\@gobble}%
\providecommand \bibinfo  [0]{\@secondoftwo}%
\providecommand \bibfield  [0]{\@secondoftwo}%
\providecommand \translation [1]{[#1]}%
\providecommand \BibitemOpen [0]{}%
\providecommand \bibitemStop [0]{}%
\providecommand \bibitemNoStop [0]{.\EOS\space}%
\providecommand \EOS [0]{\spacefactor3000\relax}%
\providecommand \BibitemShut  [1]{\csname bibitem#1\endcsname}%
\let\auto@bib@innerbib\@empty
\bibitem [{\citenamefont {Bird}(1994)}]{Bird-1994}%
  \BibitemOpen
  \bibfield  {author} {\bibinfo {author} {\bibfnamefont {G.}~\bibnamefont
  {Bird}},\ }\href@noop {} {\emph {\bibinfo {title} {Molecular Gas Dynamics and
  the Direct Simulation of Gas Flows}}}\ (\bibinfo  {publisher} {Clarendon
  Press, Oxford},\ \bibinfo {year} {1994})\BibitemShut {NoStop}%
\bibitem [{\citenamefont {Bird}(1979)}]{Bird-RGD-1979}%
  \BibitemOpen
  \bibfield  {author} {\bibinfo {author} {\bibfnamefont {G.~A.}\ \bibnamefont
  {Bird}},\ }\bibfield  {title} {\enquote {\bibinfo {title} {Simulation of
  multi-dimensional and chemically reacting flows},}\ }in\ \href@noop {} {\emph
  {\bibinfo {booktitle} {Proc.of 11th Int. Symp. on RGD}}}\ (\bibinfo {year}
  {1979})\ pp.\ \bibinfo {pages} {365--388}\BibitemShut {NoStop}%
\bibitem [{\citenamefont {Litvintsev}\ \emph {et~al.}(2024)\citenamefont
  {Litvintsev} \emph {et~al.}}]{Litvintsev-FluidDyn-2024}%
  \BibitemOpen
  \bibfield  {author} {\bibinfo {author} {\bibfnamefont {A.}~\bibnamefont
  {Litvintsev}} \emph {et~al.},\ }\bibfield  {title} {\enquote {\bibinfo
  {title} {Influence of surface chemical reactions on high-altitude
  aerothermodynamics of a generic capsule},}\ }\href {\doibase
  10.1134/S001546282460442X} {\bibfield  {journal} {\bibinfo  {journal} {Fluid
  Dynamics}\ }\textbf {\bibinfo {volume} {59}},\ \bibinfo {pages}
  {1619--–1633} (\bibinfo {year} {2024})}\BibitemShut {NoStop}%
\bibitem [{\citenamefont {Marschall}\ \emph {et~al.}(2015)\citenamefont
  {Marschall} \emph {et~al.}}]{Marschall-ChapterInBook-2015}%
  \BibitemOpen
  \bibfield  {author} {\bibinfo {author} {\bibfnamefont {J.}~\bibnamefont
  {Marschall}} \emph {et~al.},\ }\bibfield  {title} {\enquote {\bibinfo {title}
  {Surface chemistry in non-equilibrium flows},}\ }\href {\doibase
  10.2514/5.9781624103292.0239.0328} {\bibfield  {journal} {\bibinfo  {journal}
  {Hypersonic Nonequilibrium Flows: Fundamentals and Recent Advances}\ ,\
  \bibinfo {pages} {239--327}} (\bibinfo {year} {2015})}\BibitemShut {NoStop}%
\bibitem [{\citenamefont {Kroupnov}\ and\ \citenamefont
  {Pogosbekian}(2023)}]{Kroupnov-ActaAstronautica-2023}%
  \BibitemOpen
  \bibfield  {author} {\bibinfo {author} {\bibfnamefont {A.}~\bibnamefont
  {Kroupnov}}\ and\ \bibinfo {author} {\bibfnamefont {M.}~\bibnamefont
  {Pogosbekian}},\ }\bibfield  {title} {\enquote {\bibinfo {title} {Interaction
  of dissociated air with the surface of $\beta$-cristobalite material},}\
  }\href {\doibase 10.1016/j.actaastro.2022.12.027} {\bibfield  {journal}
  {\bibinfo  {journal} {Acta Astronautica}\ }\textbf {\bibinfo {volume}
  {203}},\ \bibinfo {pages} {454–--468} (\bibinfo {year} {2023})}\BibitemShut
  {NoStop}%
\bibitem [{\citenamefont {Ye}\ \emph {et~al.}(2025)\citenamefont {Ye} \emph
  {et~al.}}]{Xinbin-POF-2025}%
  \BibitemOpen
  \bibfield  {author} {\bibinfo {author} {\bibfnamefont {X.}~\bibnamefont {Ye}}
  \emph {et~al.},\ }\bibfield  {title} {\enquote {\bibinfo {title} {Chemical
  energy accommodation of atomic oxygen recombination on silica surfaces under
  high-temperature nonequilibrium flow environments: Insights from
  atomistic-scale simulations},}\ }\href {\doibase 10.1063/5.0276703}
  {\bibfield  {journal} {\bibinfo  {journal} {Physics of Fluids}\ }\textbf
  {\bibinfo {volume} {37}},\ \bibinfo {pages} {086119} (\bibinfo {year}
  {2025})}\BibitemShut {NoStop}%
\bibitem [{\citenamefont {Ivanov}\ \emph {et~al.}(2011)\citenamefont {Ivanov}
  \emph {et~al.}}]{SMILE++RGD27}%
  \BibitemOpen
  \bibfield  {author} {\bibinfo {author} {\bibfnamefont {M.}~\bibnamefont
  {Ivanov}} \emph {et~al.},\ }\bibfield  {title} {\enquote {\bibinfo {title}
  {Parallel object-oriented software system for {DSMC} modeling of
  high-altitude aerothermodynamic problems},}\ }in\ \href {\doibase
  10.1063/1.3562650} {\emph {\bibinfo {booktitle} {27th Int. Symp. on RGD}}}\
  (\bibinfo {address} {Pacific Grove, California},\ \bibinfo {year} {2011})\
  pp.\ \bibinfo {pages} {211--218}\BibitemShut {NoStop}%
\bibitem [{\citenamefont {Molchanova}, \citenamefont {Kashkovsky},\ and\
  \citenamefont {Bondar}(2014)}]{Molchanova-RGD29-2014}%
  \BibitemOpen
  \bibfield  {author} {\bibinfo {author} {\bibfnamefont {A.}~\bibnamefont
  {Molchanova}}, \bibinfo {author} {\bibfnamefont {A.}~\bibnamefont
  {Kashkovsky}}, \ and\ \bibinfo {author} {\bibfnamefont {Y.}~\bibnamefont
  {Bondar}},\ }\bibfield  {title} {\enquote {\bibinfo {title} {A detailed
  {DSMC} surface chemistry model},}\ }in\ \href {\doibase 10.1063/1.4902584}
  {\emph {\bibinfo {booktitle} {29th Int. Symp. on RGD}}}\ (\bibinfo {year}
  {2014})\ pp.\ \bibinfo {pages} {131--138}\BibitemShut {NoStop}%
\bibitem [{\citenamefont {Molchanova}, \citenamefont {Kashkovsky},\ and\
  \citenamefont {Bondar}(2018)}]{Molchanova-POF-2018}%
  \BibitemOpen
  \bibfield  {author} {\bibinfo {author} {\bibfnamefont {A.}~\bibnamefont
  {Molchanova}}, \bibinfo {author} {\bibfnamefont {A.}~\bibnamefont
  {Kashkovsky}}, \ and\ \bibinfo {author} {\bibfnamefont {Y.}~\bibnamefont
  {Bondar}},\ }\bibfield  {title} {\enquote {\bibinfo {title} {Surface
  recombination in the direct simulation {Monte Carlo} method},}\ }\href
  {\doibase 10.1063/1.5048353s} {\bibfield  {journal} {\bibinfo  {journal}
  {Physics of Fluids}\ }\textbf {\bibinfo {volume} {30}},\ \bibinfo {pages}
  {107105(18)} (\bibinfo {year} {2018})}\BibitemShut {NoStop}%
\bibitem [{\citenamefont {Shumakova}\ \emph {et~al.}(2014)\citenamefont
  {Shumakova} \emph {et~al.}}]{Shumakova-AIAA-2014}%
  \BibitemOpen
  \bibfield  {author} {\bibinfo {author} {\bibfnamefont {A.}~\bibnamefont
  {Shumakova}} \emph {et~al.},\ }\bibfield  {title} {\enquote {\bibinfo {title}
  {Effects of surface chemistry on high-altitude aerothermodynamics of space
  vehicles},}\ }\href {\doibase 10.2514/6.2014-0699} {\bibfield  {journal}
  {\bibinfo  {journal} {AIAA paper}\ ,\ \bibinfo {pages} {AIAA--2014–0699}}
  (\bibinfo {year} {2014})}\BibitemShut {NoStop}%
\bibitem [{\citenamefont {Choquet}(1999)}]{Choquet-1999}%
  \BibitemOpen
  \bibfield  {author} {\bibinfo {author} {\bibfnamefont {I.}~\bibnamefont
  {Choquet}},\ }\bibfield  {title} {\enquote {\bibinfo {title} {A new approach
  to model and simulate numerically surface chemistry in rarefied flows.}}\
  }\href {\doibase 10.1063/1.870025} {\bibfield  {journal} {\bibinfo  {journal}
  {Phys. Fluids}\ }\textbf {\bibinfo {volume} {11}},\ \bibinfo {pages}
  {1650--1661} (\bibinfo {year} {1999})}\BibitemShut {NoStop}%
\bibitem [{\citenamefont {Molchanova}(2026)}]{Molchanova-Lobachevskiy-2026}%
  \BibitemOpen
  \bibfield  {author} {\bibinfo {author} {\bibfnamefont {A.~N.}\ \bibnamefont
  {Molchanova}},\ }\bibfield  {title} {\enquote {\bibinfo {title} {Analysis of
  the truncation error of the microscopic probability function for impact
  processes in the detailed surface chemistry model},}\ }\href@noop {}
  {\bibfield  {journal} {\bibinfo  {journal} {Lobachevskii Journal of
  Mathematics}\ }\textbf {\bibinfo {volume} {47}},\ \bibinfo {pages}
  {4230--–4247} (\bibinfo {year} {2026})}\BibitemShut {NoStop}%
\bibitem [{\citenamefont {Bird}(1981)}]{Bird-AIAA-1981}%
  \BibitemOpen
  \bibfield  {author} {\bibinfo {author} {\bibfnamefont {G.}~\bibnamefont
  {Bird}},\ }\bibfield  {title} {\enquote {\bibinfo {title} {{Monte-Carlo}
  simulation in an engineering context},}\ }in\ \href {\doibase
  10.2514/5.9781600865480.0239.0255} {\emph {\bibinfo {booktitle} {Prog.
  Astronaut. Aeronaut.}}}\ (\bibinfo {year} {1981})\ pp.\ \bibinfo {pages}
  {239--255}\BibitemShut {NoStop}%
\bibitem [{\citenamefont {Gimelshein}\ and\ \citenamefont
  {Wysong}(2019)}]{Gimelshein-BirdLegacy-2019}%
  \BibitemOpen
  \bibfield  {author} {\bibinfo {author} {\bibfnamefont {S.~F.}\ \bibnamefont
  {Gimelshein}}\ and\ \bibinfo {author} {\bibfnamefont {I.~J.}\ \bibnamefont
  {Wysong}},\ }\bibfield  {title} {\enquote {\bibinfo {title} {Bird’s total
  collision energy model: 4 decades and going strong},}\ }\href {\doibase
  10.1063/1.5097706} {\bibfield  {journal} {\bibinfo  {journal} {Physics of
  Fluids}\ }\textbf {\bibinfo {volume} {31}},\ \bibinfo {pages} {076101}
  (\bibinfo {year} {2019})}\BibitemShut {NoStop}%
\bibitem [{\citenamefont {Borgnakke}\ and\ \citenamefont
  {Larsen}(1975)}]{LB-1975}%
  \BibitemOpen
  \bibfield  {author} {\bibinfo {author} {\bibfnamefont {C.}~\bibnamefont
  {Borgnakke}}\ and\ \bibinfo {author} {\bibfnamefont {S.}~\bibnamefont
  {Larsen}},\ }\bibfield  {title} {\enquote {\bibinfo {title} {Statistical
  collision model for monte carlo simulation of polyatomic gas mixture},}\
  }\href@noop {} {\bibfield  {journal} {\bibinfo  {journal} {Journal of
  Computational Physics}\ }\textbf {\bibinfo {volume} {18}},\ \bibinfo {pages}
  {405--420} (\bibinfo {year} {1975})}\BibitemShut {NoStop}%
\bibitem [{\citenamefont {Koura}\ and\ \citenamefont
  {Matsumoto}(1991)}]{Koura-PhysFluids-1991}%
  \BibitemOpen
  \bibfield  {author} {\bibinfo {author} {\bibfnamefont {K.}~\bibnamefont
  {Koura}}\ and\ \bibinfo {author} {\bibfnamefont {H.}~\bibnamefont
  {Matsumoto}},\ }\bibfield  {title} {\enquote {\bibinfo {title} {Variable soft
  sphere molecular model for inverse-power-law of lennard-jones potential},}\
  }\href {\doibase 10.1063/1.858184} {\bibfield  {journal} {\bibinfo  {journal}
  {Phys. Fluids A}\ }\textbf {\bibinfo {volume} {3}},\ \bibinfo {pages}
  {2459--2465} (\bibinfo {year} {1991})}\BibitemShut {NoStop}%
\bibitem [{\citenamefont {Kashkovsky}\ \emph {et~al.}(2025)\citenamefont
  {Kashkovsky} \emph {et~al.}}]{SMILE-family}%
  \BibitemOpen
  \bibfield  {author} {\bibinfo {author} {\bibfnamefont {A.}~\bibnamefont
  {Kashkovsky}} \emph {et~al.},\ }\bibfield  {title} {\enquote {\bibinfo
  {title} {Smile family of software systems for the direct monte carlo
  simulations of rarefied gas flows},}\ }\href {\doibase
  10.1007/s10891-026-03266-8} {\bibfield  {journal} {\bibinfo  {journal}
  {Journal of Engineering Physics and Thermophysics}\ }\textbf {\bibinfo
  {volume} {98}},\ \bibinfo {pages} {1872--1897} (\bibinfo {year}
  {2025})}\BibitemShut {NoStop}%
\bibitem [{\citenamefont {Bondar}\ \emph {et~al.}(2005)\citenamefont {Bondar}
  \emph {et~al.}}]{Accuracy-RGD24-2005}%
  \BibitemOpen
  \bibfield  {author} {\bibinfo {author} {\bibfnamefont {Y.}~\bibnamefont
  {Bondar}} \emph {et~al.},\ }\bibfield  {title} {\enquote {\bibinfo {title}
  {On the accuracy of {DSMC} modeling of rarefied flows with real gas
  effects},}\ }in\ \href {\doibase 10.1063/1.1941603} {\emph {\bibinfo
  {booktitle} {AIP Conf. Proc.}}},\ \bibinfo {series and number} {\bibinfo
  {number} {762}}\ (\bibinfo {year} {2005})\ pp.\ \bibinfo {pages}
  {607--613}\BibitemShut {NoStop}%
\bibitem [{\citenamefont {Quintas-Sánchez}\ \emph {et~al.}(2013)\citenamefont
  {Quintas-Sánchez} \emph {et~al.}}]{SurfTemp}%
  \BibitemOpen
  \bibfield  {author} {\bibinfo {author} {\bibfnamefont {E.}~\bibnamefont
  {Quintas-Sánchez}} \emph {et~al.},\ }\bibfield  {title} {\enquote {\bibinfo
  {title} {Surface temperature effects on the dynamics of n2 eley-rideal
  recombination on w(100)},}\ }\href {\doibase 10.1063/1.4774024} {\bibfield
  {journal} {\bibinfo  {journal} {The Journal of Chemical Physics}\ }\textbf
  {\bibinfo {volume} {138}},\ \bibinfo {pages} {024706} (\bibinfo {year}
  {2013})}\BibitemShut {NoStop}%
\bibitem [{\citenamefont {Verhoef}, \citenamefont {Kelly},\ and\ \citenamefont
  {Weinberg}(1993)}]{Scaling}%
  \BibitemOpen
  \bibfield  {author} {\bibinfo {author} {\bibfnamefont {R.~W.}\ \bibnamefont
  {Verhoef}}, \bibinfo {author} {\bibfnamefont {D.}~\bibnamefont {Kelly}}, \
  and\ \bibinfo {author} {\bibfnamefont {W.~H.}\ \bibnamefont {Weinberg}},\
  }\bibfield  {title} {\enquote {\bibinfo {title} {Dissociative chemisorption
  of oxygen on ir(110) as a function of angle of incidence: The effects of
  kinetic energy and surface temperature},}\ }\href {\doibase 10.1116/1.578524}
  {\bibfield  {journal} {\bibinfo  {journal} {Journal of Vacuum Science \&
  Technology A}\ }\textbf {\bibinfo {volume} {11}},\ \bibinfo {pages}
  {1926--1929} (\bibinfo {year} {1993})}\BibitemShut {NoStop}%
\bibitem [{\citenamefont {Scott}(1973)}]{Scott}%
  \BibitemOpen
  \bibfield  {author} {\bibinfo {author} {\bibfnamefont {C.}~\bibnamefont
  {Scott}},\ }\href@noop {} {\enquote {\bibinfo {title} {Wall boundary
  equations with slip and catalysis for multicomponent, nonequilibrium gas
  flows},}\ }\bibinfo {type} {Tech. Rep.}\ \bibinfo {number} {TM X-58111}\
  (\bibinfo  {institution} {NASA},\ \bibinfo {year} {1973})\BibitemShut
  {NoStop}%
\bibitem [{\citenamefont {Bonfanti}\ and\ \citenamefont
  {Martinazzo}(2016)}]{Bonfanti-QuantumChem-2016}%
  \BibitemOpen
  \bibfield  {author} {\bibinfo {author} {\bibfnamefont {M.}~\bibnamefont
  {Bonfanti}}\ and\ \bibinfo {author} {\bibfnamefont {R.}~\bibnamefont
  {Martinazzo}},\ }\bibfield  {title} {\enquote {\bibinfo {title} {Classical
  and quantum dynamics at surfaces: Basic concepts from simple models},}\
  }\href {\doibase 10.1002/qua.25192} {\bibfield  {journal} {\bibinfo
  {journal} {International Journal of Quantum Chemistry}\ }\textbf {\bibinfo
  {volume} {116}},\ \bibinfo {pages} {1575--1602} (\bibinfo {year}
  {2016})}\BibitemShut {NoStop}%
\bibitem [{\citenamefont {Gimelshein}\ \emph {et~al.}(2004)\citenamefont
  {Gimelshein} \emph {et~al.}}]{Gimelshein-PhysFluids-2004}%
  \BibitemOpen
  \bibfield  {author} {\bibinfo {author} {\bibfnamefont {S.}~\bibnamefont
  {Gimelshein}} \emph {et~al.},\ }\bibfield  {title} {\enquote {\bibinfo
  {title} {On the use of chemical reaction rates with discrete internal
  energies in the direct simulation {Monte Carlo} method},}\ }\href {\doibase
  10.1063/1.1751332} {\bibfield  {journal} {\bibinfo  {journal} {Physics of
  Fluids}\ }\textbf {\bibinfo {volume} {16}},\ \bibinfo {pages} {2442--2451}
  (\bibinfo {year} {2004})}\BibitemShut {NoStop}%
\end{thebibliography}
\providecommand{\noopsort}[1]{}\providecommand{\singleletter}[1]{#1}%

\newpage

\appendix
\begin{widetext}

\begin{center}
\textbf{Supplementary material for the paper \\ "Total Collision Energy model for surface processes in the DSMC method" \\ Ye. Bondar and A. Molchanova }
\end{center}

It is assumed that the process probability $P$ depends on $\theta$ in the following way 
\begin{equation}
P_{E,\theta} \left( E ,  \theta \right) = P(E \cos^{\chi} \theta) = P(E_{\chi}) \ , \qquad 0 \leq \chi \leq 2 \ .
\label{eq:scaling_a}
\end{equation}
This dependence interpolates between total energy scaling (TES, $\chi = 0$) and normal energy scaling (NES, $\chi=2$).

The macroscopic reaction rate is obtained by averaging $P(\Echi)$ over the
equilibrium distribution function of atoms impinging on a surface
\begin{equation}
\label{eq:integral_equation_ang_a}
K\left( T \right) = \frac{1}{n_{surf}} \sqrt{\frac{2}{\pi m}} {\left( \frac{1}{k T} \right)}^{3/2} 
\int\limits^{\pi/2}_0 \int\limits^{\infty}_0 P_{E,\theta} \left( E ,  \theta \right) \, E \exp \left( - \frac{E}{k T} \right) \, \sin \theta \cos \theta \,dE \, d\theta. \nonumber
\end{equation}
Substitute~\eqref{eq:scaling_a} and the Arrhenius dependence of the reaction rate constant into the~\eqref{eq:integral_equation_ang_a}:
\begin{equation}
\label{eq:general_equation}
A T^{b} \exp \left( - \frac{\Ea}{kT} \right)
=\frac{1}{n_{surf}} \sqrt{\frac{2}{\pi m}} (kT)^{-3/2}
\int_0^{\pi/2}\!\!\int_0^{\infty}
P\big(E\cos^{\chi}\theta\big)\,\sin\theta\cos\theta\, E\, 
\exp \left( -\frac{E}{kT} \right) \, dE \, d\theta \, .
\end{equation}

Transform the double integral on the right-hand side of the equation. 

Substitute $z = E \cos^{\chi} \theta \, , \, dz = \cos^{\chi} \theta \, dE$:
\begin{eqnarray}
\label{eq:I1}
I = 
\int_0^{\pi/2} \! \! \int_0^{\infty}
P \big( E \cos^{\chi} \theta \big) \, \sin\theta \cos\theta\, E \, 
\exp \left( -\frac{E}{kT} \right) \, dE \, d\theta \,  = \nonumber \\
= \int_0^{\pi/2} \! \! \int_0^{\infty}
z P(z) \exp \left( -\dfrac{z}{kT \cos^{\chi} \theta} \right)
\cos^{1-2\chi}\theta \ \sin \theta \, dz \, d\theta \, .
\end{eqnarray}

Perform the next change of variables $t=\cos\theta \, , \, dt=-\sin\theta\,d\theta$ in~\eqref{eq:I2}:
\begin{eqnarray}
\label{eq:I2}
I = 
\int_0^{1} \! \! \int_0^{\infty}
 \, z P(z) \, \exp \left( -\dfrac{z}{kT t^{\chi}} \right)
t^{1-2\chi} \, dz \, dt \, .
\end{eqnarray}

Introduce $s = \dfrac{1}{kT\,t^{\chi}} \, , \, t=(kTs)^{-\frac{1}{\chi}} \, , \, dt = - \frac{1}{\chi} (kT)^{-\frac{1}{\chi}} s^{-\frac{1+\chi}{\chi}} ds \,$. So:
\begin{eqnarray}
\label{eq:I3}
I = 
\frac{1}{\chi} (kT)^{\frac{2 \chi - 2}{\chi}}
\int_{\frac{1}{kT}}^{\infty} \! \! \int_0^{\infty}
\left[ \, z P(z) \, \exp \left( - s z \right) \, dz \, \right] \,
s^{\frac{\chi-2}{\chi}} \, ds \, .
\end{eqnarray}

Define the Laplace transform $\mathcal{L}\{\}$: 
\begin{equation}
\label{eq:laplace}
\mathcal{L} \{ z P(z) \} (s) = \int_0^{\infty} z\,P(z)\,e^{-sz}\,dz .
\end{equation}

Then substitute it into the~\eqref{eq:I3}:
\begin{eqnarray}
\label{eq:I4}
I = 
\frac{1}{\chi} (kT)^{\frac{2 \chi - 2}{\chi}}
\int_{\frac{1}{kT}}^{\infty} 
\mathcal{L} \{ z P(z) \} (s) \, s^{\frac{\chi-2}{\chi}} \, ds \, .
\end{eqnarray}

Denote
\begin{eqnarray}
\label{eq:G}
G(\zeta) = 
\int_{\zeta}^{\infty} 
\mathcal{L} \{ z P(z) \} (s) \, s^{\frac{\chi-2}{\chi}} \, ds \, .
\end{eqnarray}

Substitute~\eqref{eq:I4} and ~\eqref{eq:G} into the initial equation~\eqref{eq:general_equation}, collecting powers of $kT$ and replace $\frac{1}{kT} = \zeta$:
\begin{equation}
\label{eq:equation1}
A k^{-b} \zeta^{-b} \exp \left( - \frac{\Ea}{kT} \right)
=\frac{1}{n_{surf}} \frac{1}{\chi} \sqrt{\frac{2}{\pi m}} \zeta^{-\frac{1}{2}+\frac{2}{\chi}} G(\zeta) .
\end{equation}

Express the function $G(\zeta)$ from~\eqref{eq:equation1}:
\begin{equation}
\label{eq:equation2}
G(\zeta) =
A \chi n_{surf} k^{-b} \sqrt{\frac{\pi m}{2}} \zeta^{\frac{1}{2}-\frac{2}{\chi}-b} \exp \left( {- \Ea \zeta} \right) \, .
\end{equation}

Differentiate equation~\eqref{eq:equation2} with respect to the variable $\zeta$ using the rule for differentiating an integral by limit:
\begin{equation}
\label{eq:equation3}
- \zeta^{1-\frac{2}{\chi}} \, \mathcal{L} \{ z P(z) \} (\zeta) =
A \chi n_{surf} k^{-b} \sqrt{\frac{\pi m}{2}} 
\left[
\left( \frac{1}{2}-\frac{2}{\chi}-b \right) \zeta^{-\frac{1}{2}-\frac{2}{\chi}-b} - \Ea \zeta^{\frac{1}{2}-\frac{2}{\chi}-b}
\right] \, \exp \left( {- \Ea \zeta} \right)\, .
\end{equation}

Thus 
\begin{equation}
\label{eq:equation3_}
\mathcal{L} \{ z P(z) \} (\zeta) =
A \chi n_{surf} k^{-b} \sqrt{\frac{\pi m}{2}}
\left[
\Ea \zeta^{-\frac{1}{2}-b} + \left( b + \frac{2}{\chi} - \frac{1}{2} \right) \zeta^{-\frac{3}{2}-b}
\right] \, \exp \left( {- \Ea \zeta} \right)\, .
\end{equation}

Applying the inverse Laplace transform to the equation~\eqref{eq:equation3_} we obtain:
\begin{eqnarray}
\label{eq:equation4}
z P(z) && = \mathcal{L}^{-1}
\left\{
A \chi n_{surf} k^{-b} \sqrt{\frac{\pi m}{2}}
 \left[
\Ea \zeta^{-\frac{1}{2}-b} + \left( b + \frac{2}{\chi} - \frac{1}{2} \right) \zeta^{-\frac{3}{2}-b}
\right] \, \exp \left( {- \Ea \zeta} \right) \,
\right\} \, (z) \, = \nonumber \\
&& = A \chi n_{surf} k^{-b} \sqrt{\frac{\pi m}{2}}
\bigg[ 
\Ea \, \mathcal{L}^{-1} \left\{ \zeta^{-\frac{1}{2}-b} \exp \left( {- \Ea \zeta} \right) \right\} (z) + \nonumber \\
&& + \left( b + \frac{2}{\chi} - \frac{1}{2} \right) \mathcal{L}^{-1} \left\{ \zeta^{-\frac{3}{2}-b} \exp \left( {- \Ea \zeta} \right) \right\} (z)
\bigg] \, .
\end{eqnarray}

Using that
\begin{eqnarray}
\label{eq:inv_laplace_rule}
\mathcal{L}^{-1} \left\{ \zeta^{\delta} \exp \left( {- \Ea \zeta} \right) \right\} (z)
= \mathcal{L}^{-1} \left\{ \zeta^{\delta} \right\} (z-\Ea) \, \cdot \mathcal{H} (z-\Ea) = \nonumber \\
= \dfrac{(z-\Ea)^{-\delta-1}}{\Gamma(-\delta)} \cdot \mathcal{H} (z-\Ea) \, 
, -\delta > 0
\end{eqnarray}
obtain the following formula
\begin{eqnarray}
\label{eq:equation5}
z P(z)
= A \chi n_{surf} k^{-b} \sqrt{\frac{\pi m}{2}}
&& \bigg[ 
\Ea \,   \dfrac{(z-\Ea)^{-\frac{1}{2}+b}}{\Gamma(\frac{1}{2}+b)} 
+ \left( b + \frac{2}{\chi} - \frac{1}{2} \right)  \dfrac{(z-\Ea)^{\frac{1}{2}+b}}{\Gamma(\frac{3}{2}+b)} 
\bigg] \mathcal{H} (z-\Ea) \, , \nonumber \\
&& b+\frac{1}{2} > 0 \, \& \, b+\frac{3}{2} > 0  \, .
\end{eqnarray}

Express the function $P(z)$ from the equation~\eqref{eq:equation5}, combine the coefficients at $\Ea$ and $z$ and finally replace $z=\Echi$. Thus, the required formula is

\begin{eqnarray}
P(\Echi)
= \dfrac{A \chi n_{surf} k^{-b}}{\Gamma(b+\frac{3}{2})}\sqrt{\frac{\pi m}{2}}  \dfrac{(\Echi-\Ea)^{b-\frac{1}{2}}}{\Echi}
\bigg[ 
\left( b + \frac{2}{\chi} - \frac{1}{2} \right) \Echi
+\left( 1 - \frac{2}{\chi} \right) \Ea
\bigg] \mathcal{H} (\Echi-\Ea) \, , \nonumber \\
b > -\frac{1}{2} 
\end{eqnarray}

\end{widetext}

\end{document}